\documentclass{article}

\usepackage{arxiv}

\usepackage[utf8]{inputenc} 
\usepackage[T1]{fontenc}    
\usepackage{hyperref}       
\usepackage{url}            
\usepackage{booktabs}       
\usepackage{amsfonts}       
\usepackage{nicefrac}       
\usepackage{microtype}      
\usepackage{lipsum}		
\usepackage{graphicx}
\usepackage{natbib}
\usepackage{doi}

\title{An ontological analysis of misinformation in online social networks}

\author{ \href{https://orcid.org/0000-0001-7832-5081}{\includegraphics[scale=0.06]{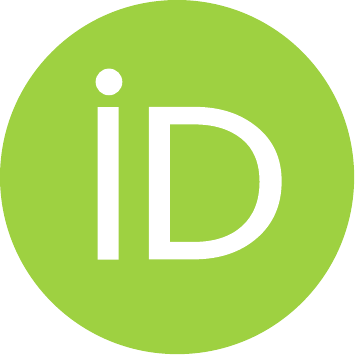}\hspace{1mm}Izzat Alsmadi}\thanks{Use footnote for providing further
		information about author (webpage, alternative
		address)---\emph{not} for acknowledging funding agencies.} \\
	Department of Computing and Cyber Security\\
	Texas A\&M, University\\
	San Antonio, TX 78259 \\
	\texttt{ialsmadi@tamusa.edu} \\
	\And
		Iyad Alazzam\\
	Department of Computer Information System\\
	Yarmouk University \\
	Irbid, Jordan \\
	\texttt{eyadh@yu.edu.jo} \\
	\And
	Mohammad A. Al-Ramahi\\
	Department of Computing and Cyber Security\\
	Texas A\&M, University\\
	San Antonio, TX 78259 \\
	\texttt{mrahman1@tamusa.edu} \\
}

\renewcommand{\shorttitle}{ An ontological analysis of misinformation in \textit{OSNs} }

\hypersetup{
	pdftitle={A template for the arxiv style},
	pdfsubject={q-bio.NC, q-bio.QM},
	pdfauthor={David S.~Hippocampus, Elias D.~Striatum},
	pdfkeywords={First keyword, Second keyword, More},
}

\begin{document}
	\maketitle
\begin{abstract}
The internet, Online Social Networks (OSNs) and smart phones enable users to create tremendous amount of information. Users who search for general or specific knowledge may not have these days problems of information scarce but misinformation. Misinformation nowadays can refer to a continuous spectrum between what can be seen as "facts" or "truth", if humans agree on the existence of such, to false information that everyone agree that it is false. In this paper, we will look at this spectrum of information/misinformation and compare between some of the major relevant concepts. While few fact-checking websites exist to evaluate news articles or some of the popular claims people exchange, nonetheless this can be seen as a little effort in the mission to tag online information with their "proper" category or label.    

\end{abstract}


\section{Introduction}
The continuous fear of the spread of misinformation risks sacrificing the values of freedom of speech as a core element in democracy. Many efforts to propose solutions to the spread of misinformation are based on enforcing some forms of censorship on who can post and what can be posted. This is already implemented in many authoritarian countries around the world who try to control public media under information censorship/accuracy claims. As one possible compromise, OSNs should encourage their users to avoid re-posting misinformation and should help them identify such misinformation. In a previous work, a model is proposed for OSNs to use and promote engagement metrics that motivate and promote positive publicity and engagement, \cite{alsmadi2016interaction}, \cite{cho2016privacy}. \\      
Facts, and truth are examples of terms that we used to refer to something that is certain or indisputable. Fact can be more often used with science, while "truth" can have belief or religion aspects. As such, what can be seen as the "truth" for someone from a certain religion, can be seen as "false truth" for someone from another religion. Religions and political orientations or beliefs, can easily make two persons disagree completely on a specific subject, news, claim, etc. This also implies an important aspect related to misinformation, is that in many cases, people unintentionally spread misinformation, thinking and believing that it is not. The statement "people hear what they want to hear" shows that in any type of information or misinformation you will find some audience who will listen to or believe in such information/misinformation. "Belief in fake news is associated with psychology, dogmatism, religious fundamentalism, and reduced analytic thinking", \cite{bronstein2019belief}.\\ 
But why these days misinformation is a big subject and concern ? What are the changes that happened in the last few years that may have triggered such issue ?

Their is no doubt that one of the main factors is the growth of Online Social Networks (OSNs) such as Twitter, Facebook, YouTube, Instagram, LinkedIn, Google+, Reddit, Snapchat, Tiktok, etc. In those websites literally all humans around the world became content generators or producers. To a large extent, their is no control on who can post and what. This is a significant change in comparison with what we had 50 or 100 years ago where governments and specific news agencies or groups control the media of TVs, newspapers, websites, etc. To a large extent, OSNs are driven by marketing and encourage users to interact more and generate more content, regardless of the credibility of such content. OSNs rely on the merchandising of a click-and-share engagement that, as a result, encourages the circulation of contents that are sticky, and “spreadable”, \cite{gerbaudo2012tweets}, \cite{venturini2019fake}. For OSN celebrities, even negative publicity can have some positive impacts, \cite{berger2010positive}.\\ 

Researches showed that people are vulnerable to the spread and exposure of misinformation because of psychological and sociological factors. Different factors such as age, political or religious orientations can impact their vulnerability, \cite{frenda2011current},  \cite{karduni2019vulnerable}.\\
In this paper we evaluated different misinformation perspectives and classifications. For example, misinformation can be broadly classified based on creator intention to intentional versus unintentional, \cite{forbes2002web}, \cite{wu2016mining}, \cite{brown2018propaganda}, \cite{rhoacutealphapitaueta2019fake}. Misinformation can also have transient and temporary effect and lasting while some other misinformation can have longer lasting and impacts. One major example of misinformation with a large impact is that which surrounded US elections 2020.
Another way to evaluate misinformation literature is to investigate actions towards misinformation, (e.g. detection, correction, prevention, etc.), \cite{chen2015deterring}. Machine learning rule is usually related to automation and replacing human efforts. Machine learning algorithms are proposed in many papers for the automatic detection of misinformation (e.g. \cite{alsmadi2019rating}, \cite{vicario2019polarization}, \cite{al2020using}, \cite{alsmadi2020toward}).

\section{Misinformation versus accurate information/facts}
The prevalence of misinformation in online social networks in several domains like science, political and health emerges new challenges. For example, propagation of rumors or misinformation versus authentic information during COVID-19 Pandemic was statistically significant \cite{tiwari2020prevalence}. Therefore, a key challenge is to come up with good mechanisms/techniques to correct misinformation on social media. Such correction mechanisms are crucial before this misinformation firmly established as accurate information and facts in receiver’s mind. Examples of those mechanisms include:
\begin{itemize}
	
\item One of these mechanisms is crowd-sourcing technique or volunteer fact-checkers that can act as social media
observers or agents \cite{vraga2017using}. In this context, \cite{vraga2019testing} explored whether observational
correction on social media by emphasizing and contradicting the logical myths in misinformation using can lead people to change their minds against controversial issues in
different domains like health, science, and politics.
\item Expert fact-checker to validate information posted \cite{collins2020fake}. Results also showed adding a
“Questioned” or “misleading” tag to misleading information on social media like false headlines makes this
information to be perceived as less accurate \cite{clayton2020real}.
 Machine learning and text mining to automatically detect and filter fake and insincere
contents in social media \cite{al2020using}.
\item Hybrid expert-machine that blends crowd and machines that showed satisfactory results in detecting fake news \cite{collins2020fake}.
\end{itemize}
Due to its sensitivity, in healthcare domain, many researchers have been attracted to examine mechanisms to identify health misinformation on social media. In this regard, 1) \cite{kim2020eye} proposed an approach based on eye tracking to accurately determine the size of attention people give to a misinformation as well as an adjustment message, and how that can be affected by the adjustment approach adopted. 2) \cite{kim2020leveraging} found that tracing replies that deliver correct information is more effective against using keywords to search for COVID-19 misinformation regarding a cure and antibiotics.

To tackle the problem of proliferation of health misinformation in online social networks, \cite{trethewey2020strategies} suggests the
following strategies:
\begin{itemize}

\item Careful dissemination of medical information: it is very important that medical research findings to be
introduced in an precise, impartial and appropriate manner so audience like news media reporters and public people can understand and act with them properly.
\item Expert-fact checking: Experts could verify tweets posted and approve them with supporting reply supplemented
with evidence.
\item Social media campaigns: there is a need to cooperate with influencers in social media such as ‘mommy bloggers’ to conduct campaigns to promote specific health information that can help spreading correct and data driven medical information to the target audiences.
\item Greater public engagement: public or expert health organizations can promote and manage campaigns of public health led by experts in different fields to engage, advice and educate public as well as highlight misinformation in different topics in their areas of expertise.
\item Fostering a fact-checking culture: it is critical to develop and encourage a philosophy of fact-checking among public and motivate them to doubt the health information communicated in social networks and see if that
information is supported by a scientific evidence.
\item Doctors as advocates: doctors and healthcare providers should be motivated to be proactive and share information supported by scientific evidence to the public through social networks outlets, like Facebook and Twitter \cite{wahbeh2020mining}.
	 
\end{itemize}

\subsection{Open issues} 
Attempts to addressing misinformation in social media need to carefully contemplate the misinformation context as well as the audience targeted in establishing efficient interventions that could be adopted by the public Vraga et al.[2019]. For example, identifying emerging health misinformation using volunteer fact checker necessitates proper context-specific keywords to acquire enough number of related potential posts and adequate advice from official healthcare persons to reduce the variation of responses Kim and Walker [2020].

\section{Misinformation versus disinformation  }
Two distinct tracks are present within popular dictionaries \cite{buckland1991information} and journalistic literature \cite{budd2011meaning} on disinformation and misinformation, for example the provenances to which designed to detect adhere. Disinformation and Misinformation may either be viewed as alternatives or differentiated in terms of meanings and deception. Misinformation can be described as accidental untruthful, imprecise or deceptive info and disinformation can be defined as untruthful, imprecise or deceptive info proposed to misinform. Within journalism, the general tendency appears designate to deal the two terms as alternatives and mostly stick to the definition of misinformation to express all kinds of fake, ambiguous, incorrect, and misleading details \cite{thorson2016belief}, \cite{wardle2018thinking}. Instead of all incorrect or inaccurate material (i.e. planned, unintentional, deceptive, deceiving, and so forth), the usage of "misinformation" underpins an appreciation of the distinction between reality and falsity between information and misinformation. Information is the real component to be maintained, covered, strengthened and disseminated. Misinformation is the incorrect component of the information to be prevented, combated, hidden and stopped. There is no difference between deliberate and deliberately deceptive and accidental mis-representative, for example honest errors, imprecision as a result of unawareness when disinformation and misinformation are regarded as synonyms. Therefore, all kinds of fraud are viewed fairly and the aim is to defend against all of them. 
It is more popular to consider misinformation and disinformation as two different terms rather than dealing with them as synonyms in the conceptual and analytical accounts of disinformation and misinformation \cite{fallis2015disinformation}, \cite{floridi2013philosophy}. In terms of motives and potential deception, the difference among disinformation and misinformation is cast: In general, misinformation is described as incorrect content, and then disinformation is described as that segment of misinformation that is in false, imprecise, or ambiguous, Notice that if disinformation is described as the deliberately deceptive component of misinformation, then in terms of motives and intention, there are no criteria for misinformation. For example, it is not possible to specify that misinformation is unintentional, Misleading as disinformation is part of Misinformation as deliberate misleading. Intentional deception should not be a subcategory of unintentional deception. Additionally, since misinformation is frequently mentioned in the sense of honest errors, prejudice, mysterious imprecision, and a distinction is maintained between misinformation and disinformation, it is fair to describe the two definitions as entirely different conceptions, wherever disinformation is not segment of misinformation \cite{burrell2016machine}.\\
 Misinformation is characterized as accidental misleading, inaccuracy, or falsehood, whereas deliberate misleading, inaccuracy, or falsehood is defined as disinformation. Intentions are also the distinctive characteristics among misinformation and disinformation: the unintentional vs the deliberate (non-accidental) misleading, imprecision, and/or falsehood \cite{capurro2003concept}. Misinformation is a false assertion that leads individuals astray by concealing the right truth. Deception, misunderstanding, falsehoods are often referred to Zhang et al. [2016]. This causes feelings of distrust that ultimately disrupt relations, which breach perceptions negatively, \cite{wu2019misinformation}, \cite{marshall2018post}.\\
  Disinformation is an incorrect part of information which is purposely circulated to confuse viewers \cite{galitsky2015detecting}. While disinformation and misinformation mutually apply to faulty or forged facts, the aim is to make a major difference between them with no intention; misinformation is applied to mislead whereas disinformation is applied with the intention \cite{kumar2016disinformation}. Disinformation, which is inaccurate or misleading information, is a subset of misinformation. It is purposely distributed to trick others online, and its effect it has continued to expand \cite{galitsky2015detecting}. In the truthful yet incorrect conviction that the spread imprecise truths remain real, misinformation is communicated. Disinformation, however, describes incorrect truths which are considered to purposefully mislead viewers and listeners. In particular, Misinformation x is misleading or incorrect knowledge that is intentionally meant to mislead. Disinformation is incorrect information designed to confuse particularly rhetoric provided to a competitor force or the press through a government department. People’s acceptance of misinformation or misleading facts depends on their past convictions and views \cite{libicki2007conquest}. The key characteristics of disinformation have been highlighted by scientists \cite{fallis2009conceptual} are: 
  
  \begin{itemize}

\item Disinformation is often the result of a deception operation that is carefully orchestrated and technically advanced. 
\item Disinformation could not come from the source who aims to mislead directly. 
\item Disinformation is frequently written to contain adjusted photographs. 
\item Disinformation may be very broadly spread or aimed at particular individuals or organizations.
\item The ultimate objective is often a person or a collection of individuals.
 
\end{itemize}

\section{Misinformation for specific event versus specific type }
Misinformation, while very popular in politics, nonetheless, can be seen in other areas such as: economics, education, culture, environments, heath and others \cite{lewandowsky2017beyond}, \cite{treen2020online}. In the U.S., in particular, 2016 election was a landmark for misinformation. Influence on the election from several actors were reported, \cite{cosentino2020polarize}, \cite{cosentino2020social}. \\
 Authors in \cite{marwick2017media} indicated that the reason behind the exploitation of fake news was a combination of political ideology and economic interests along with eager for publicity. Significant worldwide attention was given to online political campaigning through election times. OSNs such as Twitter and Facebook deleted many accounts that could be affiliated with bot/troll accounts \cite{gleicher2018election}.\\
 
 Beyond U.S. elections, many political or crisis times witnessed similar situations of large scale sharing of misinformation and possible use of bot/troll accounts. Followings are examples of such recent events:
 
 \begin{itemize}
 	\item University of Missouri protests, 2015-2016, \cite{prier2017command}, \cite{prier2017commanding}.
 	\item French elections, \cite{bulckaert2018france}, \cite{vilmer2018information}, \cite{morgan2018fake}, \cite{douglas2018religion}.
 	\item Brexit, \cite{bastos2019brexit}, \cite{narayanan2017russian}, \cite{marshall2018post}.
 	\item Anti-immigration narrative, \cite{poole2019contesting}
 	\item Estonia cyber attacks, \cite{herzog2017ten}, \cite{van2018information}.
 	\item COVID19, \cite{van2020inoculating}, \cite{apuke2021fake},  \cite{orso2020infodemic}.
 \end{itemize}

\section{Misinformation versus biased information }
Misinformation can be categorized into three different categories \cite{wani2021impact}:
\begin{itemize}
	\item It can be totally wrong.
	\item It can be a distributed perception with no genuine evidence or.
	\item It can carry misleading responses where targeted truths are provided in order to obtain some rhetoric of deception. Biased information is used when partial data is published in order to promote one side of a debate. 
	
\end{itemize}
 The study \cite{vosoughi2018spread} observed that unreliable social media information spreads much more rapidly than fact-based content. It has been noted that content truthfulness is not a motivating factor for the dissemination of information; rather, individuals prefer to spread the news based on their community favoritism, prejudice, or attention. Previous research efforts showed that the distribution of unreliable social media information has a substantial effect on terrorism \cite{oh2013community}, political campaigning \cite{bovet2019influence}, \cite{shao2018anatomy}, and crises management \cite{lukasik2016hawkes}.\\

Information bias, already a widely obvious problem in the field of media and social sciences, has also inspired a lot of empirical studies in recent years. Many media inquiries concentrate their focus on identifying information bias on specific topics such as elections, immigration, conflicts, or racism \cite{harrison2006local}.

In \cite{leban2014news} nine categories of news bias are identified:
\begin{itemize}
	\item  Bias in topic coverage
	\item Bias in speed reporting 
	\item Similarity of content 
	\item Events coverage Similarity 
	\item Geographical bias 
	\item Bias in Newswire citation 
	\item Analysis of variations in paper length 
	\item Analysis of grammatical distinctions 
	\item Variations in readability.
	
\end{itemize}

Researchers in the area of automated news bias detection often turns its emphasis to the study of sentiment and the mining of opinions in the news. Authors in \cite{leban2014news} indicated that the geographical differences/similarities between the examined news publishers are related to most forms of detected bias. For example, European news outlets put more emphasis on explaining events happening in Europe, while US publishers put more emphasis on events happening in the U.S. It is possible to find a similar trend that is related to news agency citations. Associated Press is often quoted by US news outlets, although European publishers tend to cite European news agencies. Among the tabloid outlets such as Daily Mail or Stern Magazine, they have also found a clear bias to write longer names, shorter articles and to use more descriptive language using multiple adjectives and adverbs. We also noticed a bigger percentage of adverbs and adjectives on websites, as expected.

\section{Spread of misinformation articles and fact-checking articles  }

Can we see different patterns of we see how misinformation spread versus credible information ? One problem related to misinformation correction is that in many cases the reach of a fact-check about a misinformation or claim will be less than the reach of original claim or misinformation.
It is found that in many cases, bots rather than humans spread misinformation, \cite{shao2017spread},\cite{schlitzer2018spread}, \cite{alsmadi2020many} while at the same time, the effort of human-based fact-checking websites is limited due to many factors such as the limitation or availability of expertise and resources. Some references indicate that users who spread misinformation may refer to fact-checking links that falsify such claims. Nonetheless, those users will still distribute such misinformation, \cite{shao2018anatomy}. 

Online Social Networks (OSNs) are rich platforms to spread misinformation fast as a result of the tension between aggregation of information and spread of misinformation, \cite{acemoglu2010spread}, \cite{alsmadi2019rating}, \cite{amoruso2020contrasting}. 

The type of information/misinformation has a major factor in the spread of misinformation. By far, misinformation related to politics spread much faster than any other type of misinformation. Political participation will be associated with sharing or spreading misinformation when it conforms to individuals’ beliefs, \cite{valenzuela2019paradox}. In this scope, US election in 2016 (and 2020) were milestones where the subject of misinformation in OSNs evolved rapidly.\\  

Aside from the political domain, large scale or world-wide crises such as Corona virus pandemic are usually surrounded by lots of misinformation driven by lack of credible information. For COVID-19 in particular, misinformation is related to several aspects including, virus origin, how it can reach and spread among humans, possible treatments, etc.   

Some papers investigated users based on their cognitive decisions to spread misinformation and the amount of effort they may do to fact-check such information before spreading it, \cite{greenberg2013social}, \cite{castillo2011information}, \cite{lupia2013communicating}, \cite{swire2017processing}. Researchers studied also the impact of top OSNs influencers in spreading misinformation. Misinformation distribution agents may not need to be top influencers, they could be social bots or forceful agents who emerge as dominant voices in a dispute over claims, 
\cite{acemoglu2010spread}, \cite{groshek2018media}, \cite{shao2018anatomy}.\\

The spread of misinformation can be investigated from the different possible users intentions or goals behind such act (e.g. unintentional, misleading readers, inciting clicks for revenue or supporting/manipulating
public opinions). People response to misinformation can be different and vary between:
\begin{itemize}
	\item Positive response to support and participate in spreading such misinformation.
	\item Neutral response to ignore the misinformation without any further personal analysis or response.
	\item Negative response to respond back to those who spread or originate misinformation.
\end{itemize} 

In terms of diffusion and survival, how long the spread of misinformation can survive? Can misinformation be persistent despite the fact that many fact-checking posts exist to respond to and falsify such misinformation with proof or evidence? Observations from US elections in 2016 and 2020 show that some misinformation continue to evolve and find their audience despite the response from fact-checking websites or posts. Nonetheless, researchers investigated persistence as one factor to differentiate between the spread of information versus misinformation, \cite{wang2018relevant}.  

\section{Believing versus sharing misinformation  }
What is the percentage of people who share misinformation with good intention unknowing that its not an accurate information ? If they share such misinformation intentionally what motivates them to do so? Understanding the motivations behind sharing misinformation may help us evaluating the impact as well as methods to detect and mitigate such misinformation.

Most studies assume that people do not realize the information they share is false, \cite{metzger2021dark}, \cite{talwar2019people}, \cite{duffy2020too}, \cite{chen2015students}, \cite{chadwick2019news}. Four main reasons are identified in literature behind information sharing in general, \cite{chen2015students}: entertainment, socializing, information seeking and self-expression and status seeking.   \\
We can look at those motivations from two dimensions: intentional and unintentional.

\begin{itemize}
	\item Unintentional sharing of misinformation: The following reasons can be observed in people unintentional spread of misinformation: 
	
	\begin{itemize}
		\item Sharing of information is a social activity. Users in Online Social Networks (OSNs) share their self-generated content or re-share contents from others as part of their social reach and interaction with their networks. 
		\item Publicity and social reach: People eager to share more and more information may override their eagerness and willingness to vet for information credibility. They are looking for more publicity, impact and attention. Users in OSNs receive daily large volumes of information. 
		\item Lack of resources: They may re-share information they received with their networks as part of their social interactions with little time and resources that they have to fact-check such information. Additionally, the continuous share of information among social networks makes it hard in many cases to trace and verify the content originator.  
		  
		\item Emotions: People may share misinformation when they are angry or upset as part of reaction to a crisis or negative news \cite{han2020anger}
	\end{itemize}
	
	\item Intentional sharing of misinformation: This has been associated with political or religious orientations, self-disclosure, online trust, and
	social media fatigue, \cite{talwar2019people}. People may share misinformation intentionally also to discuss it, neutralize it or correct it, \cite{rossini2020dysfunctional}.
\end{itemize} 

Political participation is positively associated with misinformation sharing, specially when it comes to misinformed users \cite{valenzuela2019paradox}, \cite{boulianne2019revolution}. Those users receive information only through their political party media or channels. While different references indicated that in the U.S. misinformation is more correlated with right-leaning partisanship, yet they do come from left party as well, \cite{nikolov2020right}. 

\section{Information/Misinformation: Binary versus multi-class }

Most research projects and applications of misinformation would like to deal with information/misinformation as a binary problem. However, in many cases, specially in OSNs context, judging information/misinformation from a binary perspective is not realistic since  in most cases, the subject content cannot be explicitly classified as absolutely true or absolutely false. Generally, they can be a combination of misinformation and true news. For such reasons, many research publications (e.g. \cite{rasool2019multi}, \cite{kaliyar2019multiclass} ) argued that a multi-class classification is more realistic.  

Another issue related to machine-based classification of misinformation is related to the different meaning and interpretation of misinformation. We see a wide range of terms used to label some kind of misinformation such as: fake news, rumors, hoaxes, insincere questions, satire, click-baits, stance, etc. In each one of those different classifications of misinformation, a machine learning algorithm will have to classify the subject content based on its specific target, regardless whether the information is correct or not. For example, in Quora classification, an insincere question can be a question that is raised not looking for an answer but rather pass a message (e.g. mocking a person, faith or culture). Additionally, a click-bait is a content in a website in which the title of that content has nothing to do with the actual content. In those examples, the machine learning algorithm will have to decide whether the content is "relevant" to the target, rather than if its correct content or not.    
\section{Summary and Conclusion}
In this paper we evaluated misinformation evolving terminologies and perspectives. Users through OSNs and smart phones can now create their own contents, respond or re-share other users' content. Each user can be literally a news channel, through their OSN page. Their is a need to redefine terms related not only to information/misinformation but also to news and news outlets. In many cases, contents generated by users will have a mixture of a piece of news (that can be correct) and users' own personal reflection or annotation. Classically, TV channels and newspapers used to be main sources of news. Nowadays news outlets are enormous and so credibility is at a serious risk.

People rely heavily on search engines to search for information. Search engines have no built-in methods to check for information credibility. Even worse, they do classify search retrieved results based on popularity rather than based on credibility-related metrics. Without building reliable methods to detect, tag and warn against inaccurate information, we will be risking civilization history.       
 
\bibliographystyle{unsrtnat}
\bibliography{references}  






\end{document}